\documentclass[aps,prl,superscriptaddress,longbibliography,twocolumn]{revtex4-2}
\usepackage{amsmath}
\usepackage{amsfonts}
\usepackage{amssymb}
\usepackage{lmodern,dsfont}
\usepackage{graphicx}
\usepackage[usenames,dvipsnames]{xcolor}
\usepackage{bm}
\usepackage[english=american]{csquotes}
\usepackage{xr}
\usepackage[caption=false]{subfig}

\newcommand{\Av}[1]{\left\langle #1 \right\rangle}
\newcommand{\av}[1]{\langle #1 \rangle}
\newcommand{\n}{\nonumber}
\newcommand{\nn}{\nonumber \\}

\newcommand{\grad}{\bm{\nabla}}

\renewcommand{\eqref}[1]{Eq.~(\ref{#1})}
\newcommand{\bmc}[1]{\bm{\mathcal{#1}}}

\begin{document}

\title{Markovian thermodynamics of non-Markovian Langevin equations}

\author{Andreas Dechant}
\affiliation{Department of Physics \#1, Graduate School of Science, Kyoto University, Kyoto 606-8502, Japan}
\author{Kiyoshi Kanazawa}
\affiliation{Department of Physics \#1, Graduate School of Science, Kyoto University, Kyoto 606-8502, Japan}

\date{\today}

\begin{abstract}
We develop the thermodynamics of non-Markovian generalized Langevin equations by embedding them in a high-dimensional Markovian representation involving auxiliary degrees of freedom.
If the memory is linear and satisfies detailed balance with the noise, we provide an explicit construction of the embedding for non-Markovian dynamics with many degrees of freedom and hydrodynamic interactions.
Moreover, while the embedding is generally not unique, we show that it results in unique values of thermodynamic quantities of the Markovian system.
This allows us to define the Markovian entropy production of a non-Markovian system, which, in contrast to the definition based directly on the non-Markovian dynamics, is guaranteed to increase monotonically with time.
Moreover, the Markovian representation allows us to identify the apparent decrease in the non-Markovian entropy with heat and information exchange between the system and the auxiliary degrees of freedom.
\end{abstract}

\maketitle

\paragraph{Introduction.}

Non-Markovian effects, where the evolution of a physical system depends on its own history, are ubiquitous in various settings, from biological cells \cite{Weber2010,Weiss2013} to quantum mechanical systems \cite{Breuer2016,Vega2017}.
They typically arise as a consequence of coarse graining \cite{Zwanzig1961,Mori1965}: If we were able to resolve all relevant environmental degrees of freedom, their present state would unambiguously determine the evolution of the system and render it Markovian.

The history-dependence of non-Markovian systems also leads to intricacies in their thermodynamic description \cite{Cockrell2022}, most prominently that they can transiently violate the second law of thermodynamics \cite{Strasberg2019}. 
Their entropy production (EP) is not a monotonic function of time and can decrease during the process, since the environmental degrees of freedom, once driven out of equilibrium, can be used to partially recover dissipated energy.

In practice, the technical complications arising from non-Markovianity are often solved by a construction called Markovian embedding \cite{Siegle2010,Goychuk2012,Ma2016,Kanazawa2020,Kanazawa2024,Kanazawa2025}, enlarging the state space of the system by adding auxiliary degrees of freedom until it can be described by Markovian evolution equations.
This can be thought of as inverse coarse graining, re-introducing artificial environmental degrees of freedom to reproduce the dynamics of the system.
However, the usefulness of this approach from a thermodynamic point of view is a more subtle question \cite{Kanazawa2025}.
Since the embedding is not unique, different Markovian representations could lead to different results for thermodynamic quantities.
This poses the question under what conditions Markovian embedding can be used to recover the thermodynamics of the environmental degrees of freedom and to resolve apparent violations of the second law.

In this Letter, we address this question a system evolving under a non-Markovian generalized Langevin equation (GLE) \cite{Zwanzig1961,Mori1965,Zwanzig1973,Fox1977,Cortes1985,DiTerlizzi2020}, which is widely used to model hydrodynamic and viscolelastic effects in artificial and biological colloidal particle systems.
We show by explicit construction how an overdamped GLE satisfying a fluctuation-dissipation relation can be embedded into an auxiliary dynamics represented by Markovian Langevin equations.
We further show that, while the embedding is not unique, it results in unique values for thermodynamic quantities such as entropy production, heat and work.
This allows us to uniquely identify a Markovian thermodynamic structure associated with the non-Markovian GLE including a positive Markovian EPR.
We also find that the apparent decrease of the non-Markovian EP can be identified with heat and information exchange between system and auxiliary degrees of freedom.
These results demonstrate that Markovian embedding can be used to characterize the thermodynamics of non-Markovian systems based on the well-established framework of Markovian stochastic thermodynamics.

\paragraph{Generalized Langevin equation.}
We suppose that the system (referred to as target system in the following) is described by a set of real-valued degrees of freedom $\bm{x}^\text{s}(t) \in \mathbb{R}^{d^\text{s}}$, which evolve according to the stochastic differential equation \cite{Zwanzig1961,Mori1965,Zwanzig1973,Fox1977,Cortes1985,DiTerlizzi2020}
\begin{align}
\int_0^t dt' \ \bm{\Gamma}(t-t') \dot{\bm{x}}^\text{s}(t') = - \grad_{x^\text{s}} U_{\bm{\lambda}(t)}(\bm{x}^\text{s}(t)) + \bm{\chi}(t) \label{gle}.
\end{align}
The left-hand side represents a non-Markovian friction force, whose value at time $t$ depends on the history of the velocity $\dot{\bm{x}}^\text{s}(t')$ at earlier times through the memory kernel $\bm{\Gamma}(t-t')$.
The friction force represents viscoelastic interactions between the target system and its environment, which is also responsible for the thermal noise $\bm{\chi}(t)$.
Since we assume that both the friction and noise originate in the environment, they satisfy a fluctuation-dissipation relation \cite{Kubo1966,Fodor2015}, that is, the statistical correlations of the noise are related to the memory kernel by
\begin{align}
\Av{\bm{\chi}(t) \bm{\chi}^\text{T}(t')} = T \bm{\Gamma}(|t-t'|), \label{fdr}
\end{align}
where $T$ is the temperature of the environment and we set the Boltzmann constant to unity.
In addition, the system is subject to forces derived from a potential $U_{\bm{\lambda}}^\text{s}(\bm{x}^\text{s})$, which may include interactions between the degrees of freedom as well as external forces; it generally depends on a set of control parameters $\bm{\lambda}$ that can be changed as a function of time.
In \eqref{gle}, we assume that all degrees of freedom $\bm{x}^\text{s}$, as well as the parameters $\bm{\lambda}$, are even under time-reversal.

Our first main result is that, if the memory kernel is of the form
\begin{align}
\bm{\Gamma}(t) = 2 \bm{\gamma}^\text{s} \delta(t) + \sum_{k=1}^{n} \bm{\gamma}_k \alpha_k e^{-\alpha_k t}, \label{memory-kernel}
\end{align}
where $\bm{\gamma}^\text{s}$ and $\bm{\gamma}_k$ are positive definite matrices and $\alpha_k > 0$ positive real numbers, then we can construct an embedding in the form of coupled Markovian Langevin equations involving (likewise even under time-reversal) auxiliary degrees of freedom $\bm{x}^\text{a}$,
\begin{subequations}
\begin{align}
\bm{\gamma}^\text{s} \dot{\bm{x}}^\text{s} &= - \grad_{x^\text{s}} U_{\bm{\lambda}}(\bm{x}^\text{s}) - \bmc{K}^\text{ss} \bm{x}^\text{s} - \bmc{K}^\text{sa} \bm{x}^\text{a} + \sqrt{2 \bm{\gamma}^\text{s} T} \bm{\xi}^\text{s}, \label{langevin-system} \\
\bm{\gamma}^\text{a} \dot{\bm{x}}^\text{a} &=  - \bmc{K}^\text{aa} \bm{x}^\text{a} - \bmc{K}^\text{as} \bm{x}^\text{s} + \sqrt{2 \bm{\gamma}^\text{a} T} \bm{\xi}^\text{a} \label{langevin-auxiliary} ,
\end{align} \label{langevin}%
\end{subequations}
where we omit the explicit dependence on $t$.
Here, $\bmc{\gamma}^a$ is symmetric and positive definite and the coupling matrix
\begin{align}
    \bmc{K} = \begin{pmatrix} \bmc{K}^\text{ss} & \bmc{K}^\text{sa} \\ \bmc{K}^\text{as} & \bmc{K}^\text{aa} \end{pmatrix}
\end{align}
is symmetric and positive semi-definite, which implies that $\bmc{K}^\text{ss}$ and $\bmc{K}^\text{aa}$ are symmetric and $\bmc{K}^\text{sa,T} = \bmc{K}^\text{as}$, where T denotes transposition.
$\bm{\xi}^\text{s}$ and $\bm{\xi}^\text{a}$ are standard Gaussian white noises with mutually independent components.
\eqref{langevin} represents a $(d = d^\text{s} + d^\text{a})$-dimensional dynamics in contact with a Markovian heat bath at temperature $T$ and governed by the potential energy $\mathcal{U}_{\bm{\lambda}}(\bm{z}) = U_{\bm{\lambda}}(\bm{x}^\text{s}) + U^\text{int}(\bm{z})$, where we refer to $\bm{z} = (\bm{x}^\text{s},\bm{x}^\text{a})$ as the extended system and $U^\text{int}(\bm{z}) = \bm{z} \cdot \bmc{K} \bm{z}/2$.
The existence of the potential is equivalent to the forces in \eqref{langevin} being conservative and follows from the symmetry of $\bmc{K}$.
We discuss the embedding procedure to obtain $\bm{\gamma}^\text{a}$ and $\bmc{K}$ form the memory kernel \eqref{memory-kernel} in detail below.
In principle, \eqref{memory-kernel} can be extended to the case $n = \infty$, which corresponds to the Laplace transform of the matrix field $\bm{\gamma}_k$ \cite{Kanazawa2020a,Kanazawa2024}.
However, the convergence of thermodynamic quantities needs to be carefully considered in this case and we restrict the following discussion to finite $n$.

Our second main result is that, while the Markovian embedding \eqref{langevin} is not unique, it is unique up to transformations that leave its thermodynamic properties invariant.
While \eqref{langevin} can easily be shown to be a sufficient condition for the memory kernel \eqref{memory-kernel} (see Appendix A), our result implies that, with regards to thermodynamics, it also necessary under the assumptions of i) linearity and ii) conservative forces.
Specifically, the EPR $\sigma_t^\text{M}$ associated with \eqref{langevin}, as well as the rate of heat $\dot{Q}_t^\text{M} = \int d\bm{z} \ \mathcal{U}_{\bm{\lambda}(t)}(\bm{z}) \partial_t p_t(\bm{z})$ exchanged between the extended system and the Markovian environment are uniquely determined by \eqref{memory-kernel} and satisfy the second law of thermodynamics \cite{Sekimoto2010,Seifert2012}
\begin{align}
\dot{S}_t^\text{M} - \frac{1}{T} \dot{Q}_t^\text{M} = \sigma_t^\text{M} \geq 0 , \label{second-law}
\end{align}
where $\dot{S}_t^\text{M}$ is the change in the differential entropy of the extended system $S_t^\text{M} = - \int d\bm{z} \ p_t(\bm{z}) \ln p_t(\bm{z})$, which is likewise invariant.
This allows us to identify the EPR $\sigma_t^\text{M}$ of the Markovian extended system as the unique Markovian EPR associated with the non-Markovian dynamics \eqref{gle}.
In particular $\sigma_t^\text{M}$ is positive, ensuring that the total entropy always increases.
By contrast, the corresponding definition \cite{Speck2007} based directly on \eqref{gle}, $\dot{S}_t^\text{s} = -d_t \int d\bm{x}^\text{s} \ p_t^\text{s}(\bm{x}^\text{s}) \ln p_t^\text{s}(\bm{x}^\text{s})$ and $\dot{Q}_t^\text{s} = \int d\bm{x}^\text{s} \ U_{\bm{\lambda}(t)}(\bm{x}^\text{s}) \partial_t p^\text{s}_t(\bm{x}^\text{s})$ does not satisfy the transient second law \cite{Strasberg2019},
\begin{align}
\dot{S}_t^\text{s} - \frac{1}{T} \dot{Q}_t^\text{s} \equiv \hat{\sigma}_t^\text{NM} \ngeq 0 \label{entropy-NM} .
\end{align}
Instead, we below derive the relation
\begin{align}
\hat{\sigma}_t^\text{NM} \geq \dot{I}_t^\text{M,s} + \frac{1}{T} \dot{Q}_t^\text{M,int},
\end{align}
where $\dot{I}_t^\text{M,s}$ is the information flow defined in Ref.~\cite{Horowitz2014} and $\dot{Q}_t^\text{M,int}$ is the heat absorbed from the auxiliary degrees of freedom.
This relation implies that an apparent decrease of EP in the non-Markovian dynamics requires either information or heat exchange between the target system and the auxiliary degrees of freedom, both of which are not captured by the non-Markovian description \eqref{gle}.

\paragraph{Markovian embedding.}
We explicitly construct a Markovian representation of \eqref{gle}, which we assume to be of the form \eqref{langevin}.
We assume that the initial state of \eqref{langevin} is
\begin{align}
p_0(\bm{z}) = p^{\text{a} \vert \text{s};\text{eq}}(\bm{x}^\text{a} \vert \bm{x}^\text{s}) p_0^\text{s}(\bm{x}^\text{s}), \label{conditional-equilibrium}
\end{align}
where $p_0^\text{s}(\bm{x}^\text{s})$ is the initial state of \eqref{gle} and $p^{\text{a} \vert \text{s};\text{eq}}(\bm{x}^\text{a} \vert \bm{x}^\text{s}) \propto \exp ( -U^\text{int}(\bm{z})/T)$ is the conditional equilibrium distribution of the auxiliary degrees of freedom.
Integrating over the auxiliary degrees of freedom, we find that the dynamics of the target system is indeed described by \eqref{gle}, see Appendix A.
More importantly, the form of \eqref{langevin} can be inferred from the memory kernel \eqref{memory-kernel} as follows.
Since $\bm{\gamma}^k$ is positive semidefinite and assuming its rank to be $r_k$, we can decompose it as
\begin{align}
\bm{\gamma}_k = \sum_{l = 1}^{r_k} \bm{c}_{k,l} \bm{c}_{k,l}^\text{T} \label{gamma-decomp}.
\end{align}
The vectors $\bm{c}_{k,l}$ are unique up to an overall sign and given $\bm{c}_{k,l} = \pm \sqrt{\beta_{k,l}} \hat{\bm{e}}_{{k,l}}$, where $\beta_{k,l} > 0$ are the non-zero eigenvalues of $\bm{\gamma}_{l,k}$ and $\hat{\bm{e}}_{l,k}$ the corresponding normalized eigenvectors.
We choose the dimension of \eqref{langevin-auxiliary} as $d^\text{a} = \sum_{k=1}^n r_k$ and define the $d^\text{a} \times d^\text{a}$ diagonal matrix $\bm{D}_{\alpha}$ that contains $r_k$ copies of each exponent $\alpha_k$ on its diagonal.
Likewise, we define the $d^\text{s}$ vector $\bm{b}_m = \bm{c}_{k,l}$, that is, the vector in \eqref{gamma-decomp} that corresponds to the $l$-th copy of $\alpha_k$.
Then, we define the coupling matrix
\begin{align}
\bmc{K} =
\begin{pmatrix}
\sum_{m=1}^{d^\text{a}} \alpha_m \bm{b}_m \bm{b}_m^\text{T} &  -\text{col}(\alpha_m \bm{b}_m) \sqrt{\bm{\gamma}^\text{a}} \\ -\sqrt{\bm{\gamma}^\text{a}} \text{row}(\alpha_m \bm{b}_m) & \sqrt{\bm{\gamma}^\text{a}} \bm{D}_{\bm{\alpha}} \sqrt{\bm{\gamma}^\text{a}}
\end{pmatrix} \label{coupling-embedding} .
\end{align}
Here, $\text{col}(\alpha_m \bm{b}_m)$/$\text{row}(\alpha_m \bm{b}_m)$ denotes a matrix whose columns/rows are the vectors $\alpha_m \bm{b}_m$, and $\bm{\gamma}^\text{a}$ is an arbitrary $d^\text{a} \times d^\text{a}$ positive definite matrix.
For this choice of the coupling matrix, the calculation in Appendix A shows that the resulting memory kernel is indeed \eqref{memory-kernel}.

The above construction provides an explicit Markovian embedding of the non-Markovian dynamics \eqref{gle}.
It also reveals that this embedding is not unique: The memory kernel and thus the GLE \eqref{gle} is invariant under arbitrary invertible transformations of the auxiliary degrees of freedom, $\tilde{\bm{x}}^\text{a} = \bm{A} \bm{x}^\text{a}$; a concrete example is a change in the friction matrix from $\bm{\gamma}^\text{a}$ to $\tilde{\bm{\gamma}}^\text{a}$, with $\bm{A} = (\sqrt{\tilde{\bm{\gamma}} ^\text{a}})^{-1} \sqrt{\bm{\gamma}^\text{a}}$.
In addition, we can arbitrarily add or remove \enquote{dummy} degrees of freedom $\bm{y}^\text{a}$, whose dynamics decouples from $\bm{x}^\text{s}$ and $\bm{x}^\text{a}$, 
\begin{align}
    \bm{\gamma}^{y,\text{a}}\dot{\bm{y}}^\text{a} = - \bmc{K}^{y,\text{aa}} \bm{y}^\text{a} + \sqrt{2 \bm{\gamma}^{y,\text{a}} T} \bm{\xi}^{y,\text{a}}, \label{langevin-dummy}
\end{align}
with positive definite matrices $\bm{\gamma}^{y,\text{a}}$ and $\bmc{K}^{y,\text{aa}}$, which obviously do not affect \eqref{gle}.
We remark that the distinction between \enquote{essential} and \enquote{dummy} auxiliary variables need not be immediately apparent in \eqref{langevin}.
However, if the memory kernel is of the form \eqref{memory-eq}, then we can always find an invertible transformation that decouples the $d^\text{a}$ variables contributing to the memory kernel from the remaining ones.

Provided that the auxiliary dynamics are linear and the coupling matrix $\bmc{K}$ is symmetric, these operations are the only ones that leave the memory kernel invariant.
A symmetric coupling matrix implies conservative forces and therefore the existence of a potential energy, with respect to which the system possesses an equilibrium state.
We remark that a similar condition was discussed in \cite{Kanazawa2025} for a thermodynamically consistent embedding of non-Markovian jump processes.
We stress that allowing for a non-symmetric coupling matrix, i.~e.~non-conservative forces, allows for more general transformations, that cannot be represented as invertible variable transformations, but leave \eqref{gle} invariant.
We discuss the resulting \enquote{non-equilibrium embedding} in detail in an upcoming publication.


\paragraph{Thermodynamic invariance.}
In Appendix B, we briefly review the thermodynamics of Markovian Langevin equations, which are discussed in-depth, e.~g.~in Refs.~\cite{Sekimoto2010,Seifert2012}.
Here, we show the invariance of various thermodynamic quantities under the transformations outlined above.
We write an invertible linear transformation $\bm{A}$ of the auxiliary degrees of freedom as $\tilde{\bm{z}} = \bmc{A} \bm{z}$, where the matrix $\bmc{A}$ acts as the identity on $\bm{x}^\text{s}$ and as $\bm{A}$ on $\bm{x}^\text{a}$.
In the Langevin equation \eqref{langevin}, this results in a transformed friction and coupling matrix,
\begin{align}
    \tilde{\bm{\gamma}} = \bmc{A}^{\text{T},-1} \bm{\gamma} \bmc{A}^{-1}, \qquad \tilde{\bmc{K}} = \bmc{A}^{\text{T},-1} \bmc{K} \bmc{A}^{-1} ,
\end{align}
where $\bm{\gamma}$ is the block-diagonal matrix with entries $\bm{\gamma}^\text{s}$ and $\bm{\gamma}^\text{a}$.
In terms of the Fokker-Planck equation \eqref{fpe}, this implies that the local mean velocity transforms as
\begin{align}
    \tilde{\bm{\nu}}_t(\tilde{\bm{z}}) = \bmc{A} \bm{\nu}_t \big(\bmc{A}^{-1} \tilde{\bm{z}} \big) ,
\end{align}
and, therefore, we have for the Markovian EPR \eqref{entropy-markov},
\begin{align}
    \tilde{\sigma}_t^\text{M} = \sigma_t^\text{M}.
\end{align}
The invariance of EPR holds not only for linear but for arbitrary invertible variable transformations.
While the differential entropy is not invariant under linear variable transformations,
\begin{align}
    \tilde{S}_t^\text{M} = S_t^\text{M} - \ln \vert \det(\bm{A}) \vert,
\end{align}
it only changes by an additive constant and, thus, $\dot{S}_t^\text{M}$ is invariant.
From \eqref{second-law}, this implies that the heat flow $\dot{Q}_t^\text{M}$ is likewise invariant.

On the other hand, if the initial state of the auxiliary degrees of freedom is in conditional equilibrium, then the dummy degrees of freedom \eqref{langevin-dummy} remain in equilibrium independent of the protocol acting on the target system and thus do not contribute to either EPR or entropy change.
Thus, the Markovian change in heat, entropy and free energy, as well as the EP are all uniquely determined.
Remarkably, while the auxiliary degrees of freedom are not unique, their thermodynamics is completely determined by \eqref{gle}, under the assumptions that the forces are i) linear and ii) conservative.

\paragraph{Non-Markovian entropy production.}
We now compare the Markovian EPR \eqref{entropy-markov} to its non-Markovian counterpart defined in \eqref{entropy-NM}.
In contrast to $\sigma_t^\text{M}$, the non-Markovian expression $\hat{\sigma}_t^\text{NM}$ is not always positive and therefore an apparent EPR \cite{Strasberg2019}.
This apparent non-positivity of EPR is reminiscent of information thermodynamics \cite{Sagawa2012,Ito2013,Hartich2014,Horowitz2014,Parrondo2015}, which we briefly review in Appendix C.
In particular, we can assign part of the Markovian heat flow to the target system (see \eqref{heat-decomposition}),
\begin{align}
    \dot{Q}_t^\text{M,s} = \Av{\grad_{x^\text{s}} \mathcal{U}_{\bm{\lambda}} \circ \dot{\bm{x}}^\text{s}} .
\end{align}
This defines the apparent EPR of the target system in the Markovian picture, which satisfies the second law of information thermodynamics,
\begin{align}
    \hat{\sigma}_t^\text{M,s} = \dot{S}_t^\text{s} - \frac{1}{T} \dot{Q}_t^\text{M,s} \geq \dot{I}_t^\text{M,s}  \label{second-law-info}
\end{align}
with the information flow $\dot{I}_t^\text{M,s}$ \cite{Horowitz2014}, see \eqref{information-flow}.
Recalling that $\mathcal{U}_{\bm{\lambda}}(\bm{z}) = U_{\bm{\lambda}}(\bm{x}^\text{s}) + U^\text{int}(\bm{z})$, we have
\begin{align}
\dot{Q}_t^\text{M,s} = \dot{Q}_t^\text{s} + \dot{Q}_t^\text{M,int} \quad \text{with} \quad \dot{Q}_t^\text{M,int} = \Av{\grad_{x^\text{s}} U^\text{int} \circ \dot{\bm{x}}^\text{s}}, \label{interaction-heat}
\end{align}
where $\dot{Q}_t^\text{M,int}$ is the heat flow into the target system due to interactions with the auxiliary degrees of freedom.
Comparing \eqref{entropy-NM} and \eqref{second-law-info}, we then obtain for the non-Markovian EPR
\begin{gather}
\hat{\sigma}^\text{NM}_t \geq \dot{I}_t^\text{M,s} + \frac{1}{T} \dot{Q}_t^\text{M,int}  \label{entropy-non-markov-information} .
\end{gather}
Thus, a decrease in the non-Markovian EP can occur for two distinct reasons:
The first, information-thermodynamic contribution is a negative information flow, similar to \eqref{second-law-info}.
In contrast to the second law of information thermodynamics, however, there exists a second, energetic contribution due to heat absorbed from the auxiliary degrees of freedom.
In particular, the non-Markovian EP can decrease even in the absence of information flow.
We remark that, just like the overall EPR and heat flow, both the information flow $\dot{I}_t^\text{M,s}$ and the interaction heat flow $\dot{Q}_t^\text{M,int}$ are independent of the choice of the embedding variables (see Appendix C) and thus uniquely defined by \eqref{gle}.

\begin{figure*}
\parbox{0.53\textwidth}{\includegraphics[width=0.53\textwidth]{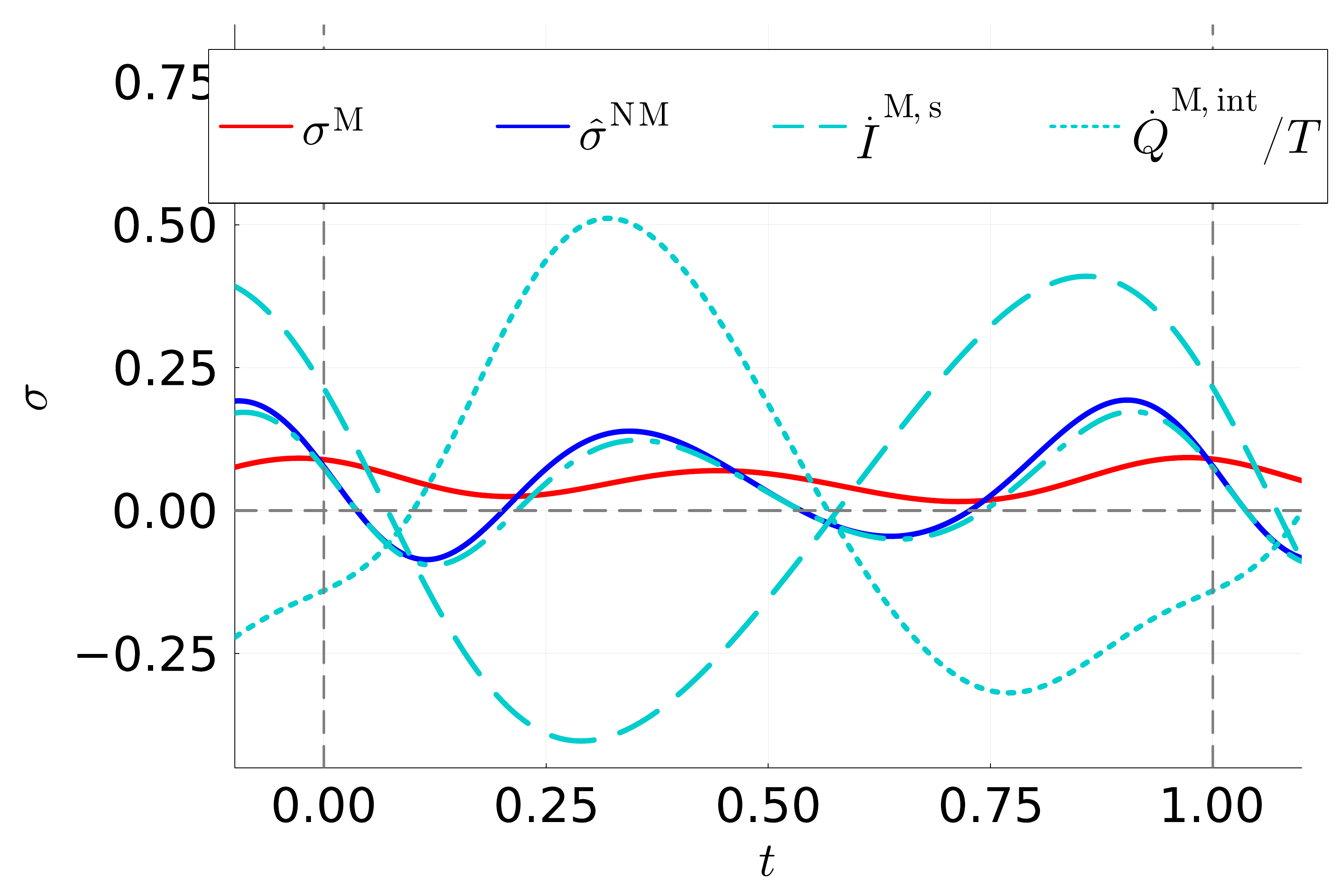}}
\parbox{0.45\textwidth}{\subfloat{\includegraphics[width=0.45\textwidth]{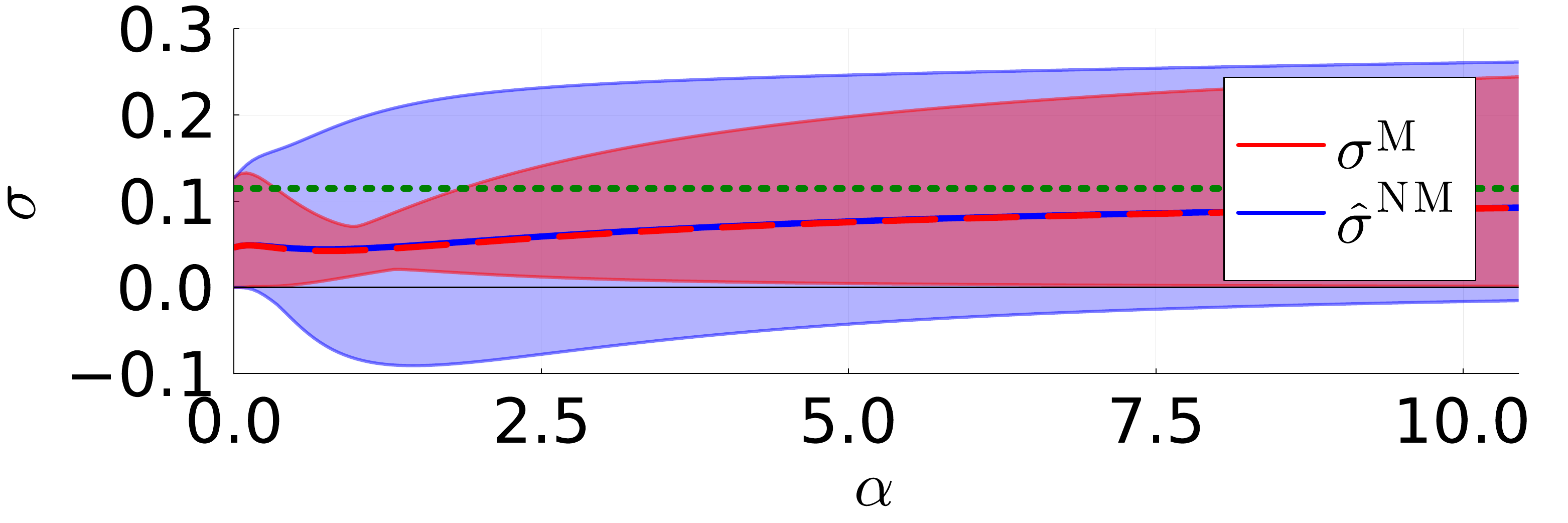}}\\
\subfloat{\includegraphics[width=0.45\textwidth]{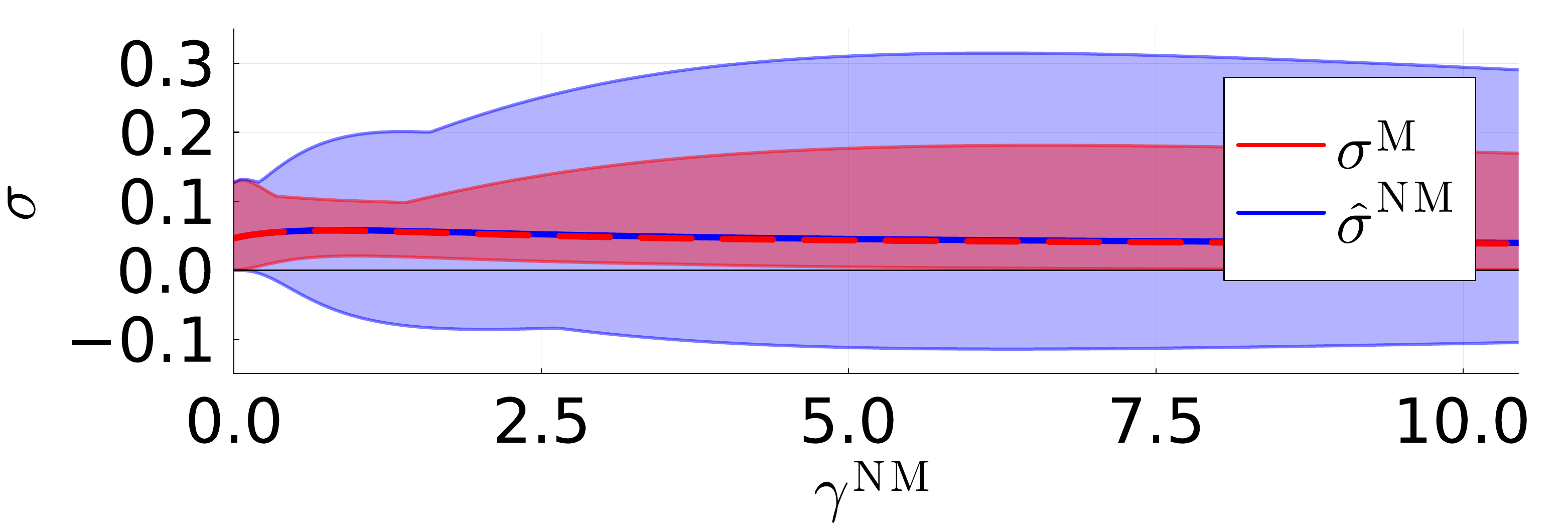}}}
\caption{Markovian and non-Markovian EPR for a particle in a trap with periodically varying trap stiffness. 
Left: As a function of time for one period of the driving (indicated by the dashed vertical lines). 
The red line is the Markovian EPR \eqref{second-law}, the blue line the apparent non-Markovian EPR \eqref{entropy-NM}. 
The light blue lines are the information flow \eqref{information-flow} (dashed), the heat flow due to the interaction with the auxiliary mode \eqref{interaction-heat} (dotted) and the sum of both terms (dash-dotted). 
Right: As a function of the decay rate (top) and magnitude (bottom) of the memory friction. The solid lines are the time-averaged EPRs, the shaded area indicates the range of the EPRs over one period of the driving. 
Parameters are (except otherwise noted) $\lambda_0 = 2$, $\Delta\lambda = 1$, $\omega = 2 \pi$, $\gamma^\text{s} = 0.05$, $\alpha = 2$, $\gamma^\text{NM} = 2$. 
The green dashed line in the top panel indicates the time-averaged EPR for a Markovian system with $\gamma^\text{s} = 2.05$.}
\label{fig_trap_stiffness}
\end{figure*}

\paragraph{Heat and information flow.}
To illustrate the above results, we first consider a simple one-dimensional system with memory kernel
\begin{align}
\Gamma(t) = 2 \gamma^\text{s} \delta(t) + \gamma^\text{NM} \alpha e^{-\alpha t},
\end{align}
corresponding to a single exponential mode with decay rate $\alpha$.
We set $\gamma^\text{s} = 0.05$, so that the system is dominated by the non-Markovian friction.
We drive the system by varying the trap stiffness of a parabolic trap $U_{\lambda(t)}(x^\text{s}) = \lambda(t)(x^\text{s})^2/2$ periodically in time as $\lambda(t) = \lambda_0 + \Delta\lambda \sin(\omega t)$.
Since the system is linear, we can compute all thermodynamic quantities from the mean and covariance matrix, see Appendix D.
The left panel of Fig.~\ref{fig_trap_stiffness} shows EPR as a function of time in the periodic steady state.
While the Markovian EPR is always positive, we clearly see that the apparent non-Markovian EPR takes negative values during some parts of the driving protocol.
The non-Markovian EPR is lower bounded by the sum of information and interaction heat flow according to \eqref{entropy-non-markov-information}, however, it can be negative if either the information flow or the interaction heat flow are positive, indicating that neither effect alone is sufficient to explain the occurrence of apparent negative EPR.

The right panel of Fig.~\ref{fig_trap_stiffness} shows EPR as a function of the memory parameters $\alpha$ (top) and $\gamma^\text{NM}$ (bottom).
The Markovian and non-Markovian EPRs agree when averaged over one period of the driving (solid lines). 
We remark that this is not a coincidence; indeed the time-integrated Markovian and non-Markovian EPs can be shown to only differ by a contribution that is non-extensive in time \cite{Funo2026}.
However, the respective range over which the EPRs vary during one period (shaded areas) are markedly different and only agree in the limits where the influence of the non-Markovian friction disappears.
In particular, even when the temporal variation of the Markovian EPR is small, the non-Markovian EPR can vary significantly and take large negative values.
This suggests that, while non-Markovian EP is a useful measure of accumulated EP during a process, it fails to correctly capture the transient behavior of dissipation, in particular when the characteristic timescales of the memory are comparable to those of the driving.
When comparing the EPR with a Markovian dynamics with the same overall friction coefficient $\gamma^\text{M} = \gamma^\text{s} + \gamma^\text{NM}$ (dashed green line in the top-right panel of Fig.~\ref{fig_trap_stiffness}), we see that the EPR slowly converges to the Markovian limit as $\alpha \rightarrow \infty$. 

\begin{figure*}
\includegraphics[width=0.47\textwidth]{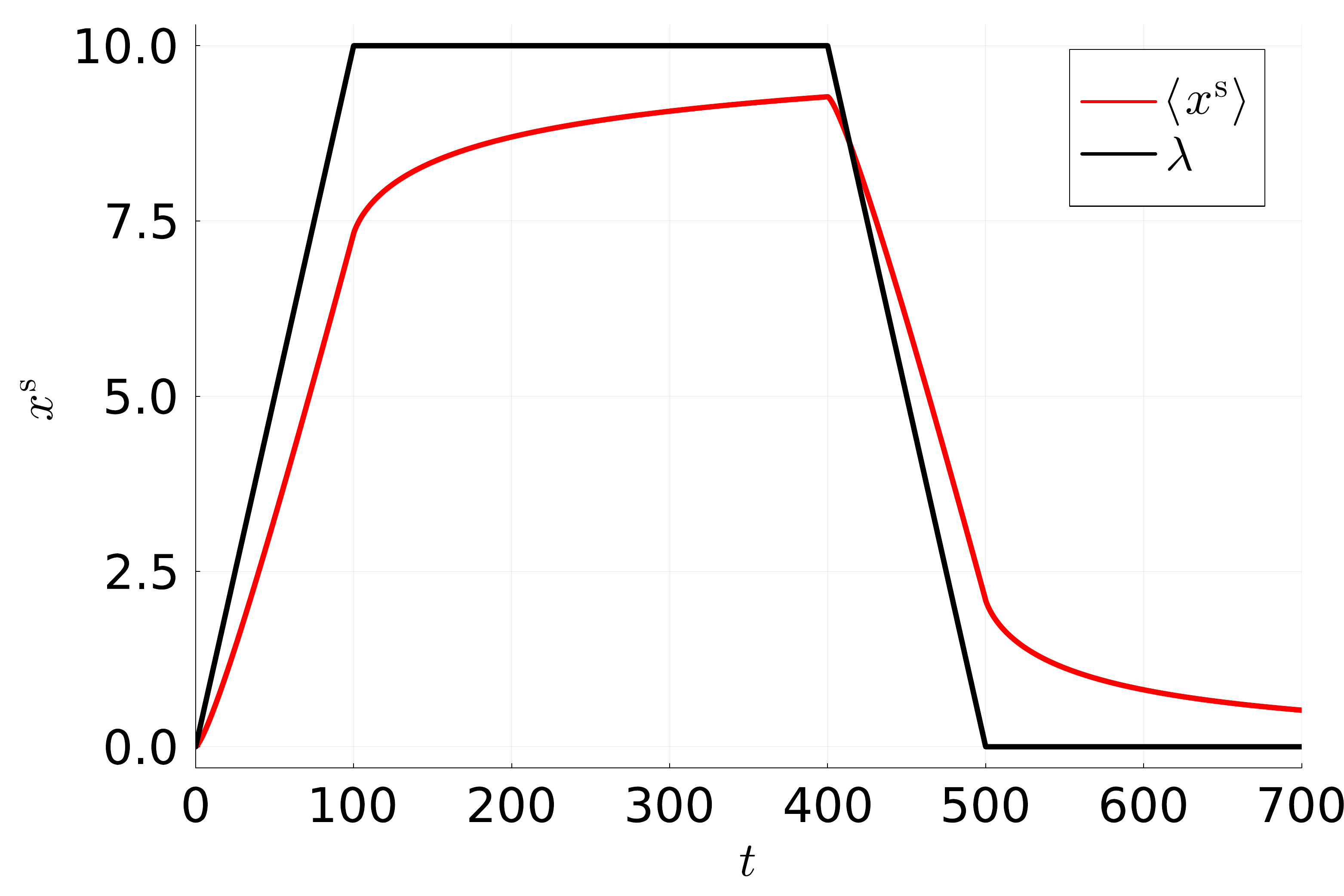}
\includegraphics[width=0.47\textwidth]{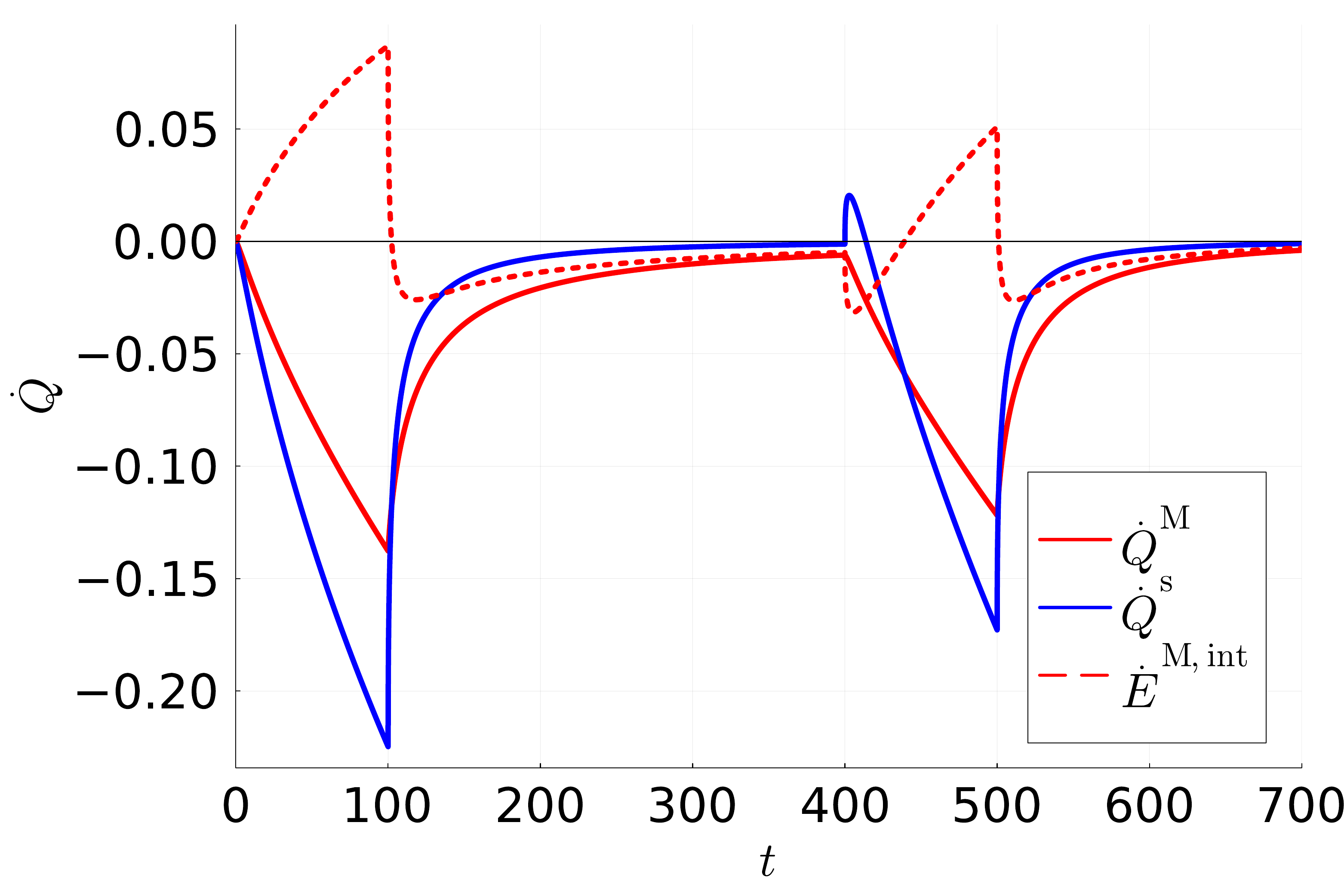}
\caption{Heat flow for a driven particle with power-law memory.
Left: Position of the trap $\lambda(t)$ (purple) and the resulting mean position of the particle $\av{x^\text{s}(t)}$ (red).
Right: Markovian heat flow in \eqref{second-law} (red) and non-Markovian heat flow in \eqref{entropy-NM} (blue), as well as the rate of change of interaction energy $\dot{E}^\text{M,int} = d_t \Av{U^\text{int}}$ (red dashed).
Parameters are $\gamma^\text{s} = 0.1$, $\gamma^\text{NM} = 100$, $k^\text{s} = 1$, while the power-law memory is given by \eqref{memory-power-law} with $\tau_0 = 0.01$, $\tau_1 = 1000$ and $\beta = 0.5$, and we use $d^\text{a} = 8$ exponential modes to fit it.}
\label{fig_power_law}
\end{figure*}

\paragraph{Energetics of long-range memory.}
The same approach can also accommodate long-ranged power-law memory which is associated with subdiffusive motion \cite{Lutz2001,Deng2009} and often encountered in viscoelastic biological systems \cite{Weber2010,Weiss2013}.
Concretely, we consider a memory kernel of the form
\begin{align}
    \Gamma(t) = 2 \gamma^\text{s} \delta(t) + \gamma^\text{NM} \frac{\big(1+\frac{t}{\tau_0} \big)^{-\beta} e^{-\frac{t}{\tau_1}}}{\tau_0 e^{\tau_0/\tau_1} \text{E}_\beta(\tau_0/\tau_1)} \label{memory-power-law} ,
\end{align}
with $\beta > 0$.
This describes a memory that decays as a power law $t^{-\beta}$ for $\tau_0 < t < \tau_1$, while being constant for $t < \tau_0$ and decaying exponentially for $t > \tau_1$. 
Here, $E_\beta(x)$ is the exponential integral function, which normalizes the memory kernel such that $\int_0^\infty dt \ \Gamma(t) = \gamma^\text{s} + \gamma^\text{NM}$.
It has been shown previously \cite{Siegle2010,Siegle2011,Goychuk2012} that a power-law memory kernel can be well approximated by a even a small number of exponentials.
Concretely, we approximate \eqref{memory-power-law} with \eqref{memory-kernel} and a given number $d^\text{a}$ of modes, determining the coefficients $c_k = b_k^2$ and exponents $\alpha_k$ by fitting.
We consider a particle in a parabolic trap and vary the trap position as a function of time, $U_{\lambda(t)}(x^\text{s}) = k^\text{s}(x^\text{s} - \lambda(t))^2/2$.
We remark that, for linear dynamics, a change in the average position does not change the Shannon entropy; consequently, EPR is determined solely by the heat flow.

The resulting average position of the particle is shown in the left panel of Fig.~\ref{fig_power_law}.
We observe that, while the trap is moving, the long-range memory leads to an increasing lag between the trap position and the particle.
At the same time, both the Markovian and non-Markovian heat flow (red and blue line, right panel of Fig.~\ref{fig_power_law}) become increasingly negative, corresponding to an increasing rate of heat dissipation.
Once the trap stops moving, the particle position relaxes towards the new equilibrium value.
During the relaxation, the magnitude of the non-Markovian heat flow decreases rapidly and almost vanishes, suggesting no heat being dissipated and the target system being close to equilibrium.
Surprisingly, when we reverse the motion of the trap, we observe a positive non-Markovian heat flow and thus negative dissipation starting from an apparent near-equilibrium state.
By contrast, in the Markovian description, we see that the out-of-equilibrium auxiliary degrees of freedom supply energy to the target system, explaining the apparent negative dissipation.
We remark that these results are insensitive to the details of the embedding, in particular, they rapidly converge as the number of auxiliary degrees of freedom $d^\text{a}$ increases.
Doubling the number of modes from 8 to 16 leads to no discernible difference in Fig.~\ref{fig_power_law}.


\paragraph{Discussion.}
The results obtained in this Letter imply that a non-Markovian GLE \eqref{gle} possesses a unique Markovian thermodynamic structure, in which thermodynamic quantities do not depend on the details of the Markovian representation.
Under the hypothesis that non-Markovian dynamics always represent an effective description of an underlying Markovian system, we argue that the Markovian entropy production is more fundamental, since it explicitly accounts for the transient non-equilibrium state of the environment and resolves apparent negative entropy production rates \cite{Strasberg2019}.
It would be interesting to investigate whether a similar thermodynamic interpretation can be given to other Markovian representations of non-Markovian processes, such as in the case of weak memory \cite{Brandner2025}.

Our general framework has several prospective applications:
While the auxiliary dynamics and the coupling to the latter are assumed as linear, the forces acting on the target system can be non-linear.
This allows extending models of information erasure that represent a bit of information via the minima of a potential \cite{Berut2012,Proesmans2020} to the non-Markovian regime.
For such non-monotonic protocols, we expect significant differences between the Markovian and non-Markovian EP in finite-time erasure.
Recently, it was found that minimum-work protocols in systems with memory rely on partial recovery of dissipated work \cite{Loos2024,Monter2025}.
Our framework uniquely characterizes both energy and information exchange with the environment and can elucidate their role in optimal protocols.
Finally, we anticipate an application to processes in viscoelastic environments, for example when moving a tracer particle with an optical tweezer through the cytoskeleton of biological cells \cite{Weber2010,Weiss2013}.
Here, our framework would allow to reconstruct the energy transfer to and dissipation of the cytoskeleton based solely on the experimentally observed motion of the tracer particle and its effective description via the GLE.

Finally, let us remark on the connection between the present work and our earlier work \cite{Kanazawa2025}, which focuses on the complementary class of non-Markovian jump processes.
Intriguingly, even though the concrete dynamics differ considerably, in both cases, we can recover a Markovian thermodynamic structure that is invariant under the embedding under suitable conditions.
This, together with other recent studies \cite{Funo2026}, suggests that, on the one hand, Markovian embedding can become a useful tool for accurately characterizing the thermodynamics of finite-time processes in non-Markovian settings. 
On the other hand, it also stresses the importance of identifying the conditions under which entropy production is invariant under the representation of the observed dynamics.

\textbf{Acknowledgments:}
AD was supported by JSPS KAKENHI (Grants No. 22K13974, 24H00833 and 25K00926).
KK was suported by JSPS KAKENHI (Grant No. 23H00467 and 25K24774).
This work was supported by JST ERATO Grant Number JPMJER2302, Japan.
AD gratefully acknowledges discussions with Ken Funo, Jann van der Meer and Kay Brandner.

\bibliography{JabRef_main.bib,localbib.bib}

\pagebreak
\clearpage

\section{End matter}
\appendix

\paragraph{Appendix A: Derivation of the generalized Langevin equation.}
We note that the form of the memory kernel \eqref{memory-general} implies that the noises acting on the target and auxiliary system in \eqref{langevin} are independent; without this assumption the control parameters $\bm{\lambda}(t)$ appear explicitly in the memory kernel.
The derivation of \eqref{gle} from \eqref{langevin} can be found in various references, e.g.~\cite{Zwanzig1973,Fox1977,Cortes1985}, but we provide it in our notation for convenience.
The formal solution of \eqref{langevin-auxiliary} is
\begin{align}
    \bm{x}^\text{a}(t) &= e^{-\bmc{M}^\text{aa} t} \bm{x}^\text{a}(0) + \int_0^t dt' \ e^{-\bmc{M}^\text{aa} (t-t')} \nn
    &\times \Big( -\bmc{M}^\text{as} \bm{x}^\text{s}(t') + \sqrt{2 (\bm{\gamma}^\text{a})^{-1} T} \bm{\xi}^\text{a}(t') \Big) ,
\end{align}
defining $\bmc{M}^\text{aa} = (\bm{\gamma}^\text{a})^{-1} \bmc{K}^\text{aa}$, $\bmc{M}^\text{as} = (\bm{\gamma}^\text{a})^{-1} \bmc{K}^\text{as}$ and $\bmc{M}^\text{sa} =  \bmc{K}^\text{sa} (\bm{\gamma}^\text{a})^{-1}$ for convenience of notation.
Plugging this into \eqref{langevin-system} and integrating by parts with respect to time, we obtain
\begin{align}
    \int_0^t dt' &\ \bm{\Gamma}(t-t') \dot{\bm{x}}^\text{s}(t') \\
    &= - \grad_{x^\text{s}} U_{\bm{\lambda}(t)}(\bm{x}^\text{s}(t)) - \bm{K}^\text{eff} \bm{x}^\text{s}(t) + \bm{\chi}(t) \n,
\end{align}
where we defined
\begin{subequations}
\begin{align}
    \bm{\Gamma}(t) &= 2 \bm{\gamma}^\text{s} \delta(t) + \bmc{M}^\text{sa} \bm{\gamma}^\text{a} (\bmc{M}^\text{aa})^{-1} e^{-\bmc{M}^\text{aa} t} \bmc{M}^\text{as}, \label{memory-general} \\
    \bm{K}^\text{eff} &= \bmc{K}^\text{ss} - \bmc{K}^\text{sa} (\bmc{K}^\text{aa})^{-1} \bmc{K}^\text{as} \\
    \bm{\chi}(t) &= \sqrt{2 \bm{\gamma}^\text{s} T} \bm{\xi}^\text{s}(t), \\
    &\quad- \bmc{K}^\text{sa} \int_0^t dt' \ e^{-\bmc{M}^\text{aa} (t-t')} \sqrt{2 (\bm{\gamma}^\text{a})^{-1} T} \bm{\xi}^\text{a}(t') \nn
    &\quad+ \bmc{K}^\text{sa} e^{-\bmc{M}^\text{aa} t} \big( \bm{x}^\text{a}(0) - (\bmc{M}^\text{aa})^{-1} \bmc{M}^\text{as} \bm{x}^\text{s}(0) \big) \n .
\end{align}
\end{subequations}
Due to the assumption that $\bmc{K}$ is symmetric with $\bmc{K}^\text{ss} = \bmc{K}^\text{sa} (\bmc{K}^\text{aa})^{-1} \bmc{K}^\text{as}$, we have $\bm{K}^\text{eff} = 0$, which means that the auxiliary degrees of freedom do not lead to any systematic force on the target system.
The noise $\bm{\chi}(t)$ generally has non-stationary correlations and depends on the initial configuration of both the target and auxiliary system.
However, under the assumption that the conditional equilibrium distribution of the auxiliary degrees of freedom is in equilibrium, \eqref{conditional-equilibrium}, the statistics of the noise conditioned on the initial condition $\bm{x}^\text{s}(0)$ are Gaussian and determined by
\begin{subequations}
\begin{align}
    \Av{\bm{\chi}(t) \vert \bm{x}^\text{s}(0)} &= 0, \\ 
    \Av{\bm{\chi}(t) \bm{\chi}^\text{T}(t') \vert \bm{x}^\text{s}(0)} &= T \bm{\Gamma}(|t-t'|) ,
\end{align}
\end{subequations}
which is independent of $\bm{x}^\text{s}(0)$, resulting in \eqref{gle}.
We remark that, in contrast to the Markovian limit, where the auxiliary degrees of freedom relax to equilibrium almost instantaneously, in the absence of such a time scale separation, the condition of conditional equilibrium is generically satisfied only when the entire system is initially in equilibrium.

To obtain \eqref{memory-kernel}, we note that, since $\bmc{K}^\text{aa}$ is symmetric and positive definite, so is the matrix $\bmc{Q}^\text{aa} =(\sqrt{\bm{\gamma}^\text{a}})^{-1} \bmc{K}^\text{aa} (\sqrt{\bm{\gamma}^\text{a}})^{-1}$ and we can diagonalize the latter by an orthogonal transformation.
Moreover, we have $\bmc{M}^\text{aa} = (\sqrt{\bm{\gamma}^\text{a}}) \bmc{Q}^\text{aa} (\sqrt{\bm{\gamma}^\text{a}})^{-1}$, which is similar to $\bmc{Q}^\text{aa}$ and shares the same positive eigenvalues.
Introducing the orthogonal matrix $\bm{R}$ diagonalizing $\bmc{Q}^\text{aa}$ and the diagonal matrix $\bm{D}_\alpha$,
\begin{align}
\bm{R}^\text{T} \bmc{Q}^\text{aa} \bm{R} = \bm{D}_\alpha,
\end{align}
we obtain from \eqref{memory-general}
\begin{align}
\bm{\Gamma}(t) = 2\bm{\gamma}^\text{s} \delta(t) + \sum_{k=1}^{d^\text{a}} \bm{b}_k \bm{b}_k^\text{T} \alpha_k e^{-\alpha_k t}. \label{memory-eq}
\end{align}
Here, $\alpha_k > 0$ are the eigenvalues of $\bmc{M}^\text{aa}$ and $\bm{b}_k$ are coefficient vectors given by
\begin{align}
\bm{b}_{k} = \frac{1}{\alpha_k}\big( \bmc{K}^\text{sa} (\sqrt{\bm{\gamma^\text{a}}})^{-1} \bm{R} \big)_{\bullet k}, \label{coefficient-vector}
\end{align}
where $(\bm{A})_{\bullet k}$ denotes the $k$-th column of a matrix $\bm{A}$.
\eqref{memory-eq} is a superposition of exponentially decaying modes with decay rates $\alpha_k$, weighted by the positive semidefinite rank-1 matrices $\bm{\gamma}_k = \bm{b}_k \bm{b}_k^\text{T}$.

\paragraph{Appendix B: Thermodynamics of Markovian Langevin equations.} \label{app_thermo}
The thermodynamic description of Markovian Langevin equations is well-established \cite{Sekimoto2010,Seifert2012} and we only briefly summarize the results.
The Markovian dynamics \eqref{langevin} is equivalent to the Fokker-Planck equation for the probability density $p_t(\bm{z})$,
\begin{align}
\partial_t p_t(\bm{z}) &= - \grad_z \cdot \big( \bm{\nu}_t(\bm{z}) p_t(\bm{z}) \big) \quad \text{with} \quad \label{fpe} \\
\bm{\nu}_t(\bm{z}) &= - \bm{\gamma}^{-1} \big( \grad_z \mathcal{U}_{\bm{\lambda}(t)}(\bm{z}) + T \grad \ln p_t(\bm{z}) \big) \n ,
\end{align}
where $\bm{\gamma}$ is the block-diagonal matrix with entries $\bm{\gamma}^\text{s}$ and $\bm{\gamma}^\text{a}$.
The quantity $\bm{\nu}_t(\bm{z})$ is called the local mean velocity; it represents the velocity of probability flows in the system and determines the EPR
\begin{align}
\sigma_t^\text{M} = \frac{1}{T} \Av{\bm{\nu}_t \cdot \bm{\gamma} \bm{\nu}_t}_t \label{entropy-markov},
\end{align}
where $\av{\ldots}_t$ denotes an average with respect to $p_t(\bm{z})$ and we use the superscript M to denote quantities that are computed using the Markovian dynamics \eqref{langevin}.
We note that $\sigma_t^\text{M} \geq 0$ with equality if and only if $p_t(\bm{z}) = p_{\bm{\lambda}}^\text{eq}(\bm{z}) \propto e^{-\mathcal{U}_{\bm{\lambda}}(\bm{z})/T}$, that is, when the extended system is in equilibrium.
We define the internal energy of the extended system as the mean value of the potential $E^\text{M}_t = \av{\mathcal{U}_{\bm{\lambda}(t)}}_t$.
Its time-derivative yields the first law 
\begin{align}
\dot{E}^\text{M}_t = \dot{W}^\text{M}_t + \dot{Q}^\text{M}_t, \label{first-law}
\end{align}
where the rate of work $\dot{W}^\text{M}_t$ and heat $\dot{Q}^\text{M}_t$ transferred to the system from the environment are defined as \cite{Sekimoto2010,Seifert2012}
\begin{subequations}
\begin{align}
\dot{W}^\text{M}_t &= \Av{\grad_{\lambda} \mathcal{U}_{\bm{\lambda}}}_t \cdot \dot{\bm{\lambda}} = \Av{\grad_\lambda U_{\bm{\lambda}}}_t^\text{s} \cdot \dot{\bm{\lambda}} = \dot{W}_t^\text{s}, \\
\dot{Q}^\text{M}_t &= \Av{\grad_z \mathcal{U}_{\bm{\lambda}(t)} \circ \dot{\bm{z}}} ,
\end{align} \label{heat-work-markov}%
\end{subequations}
where $\circ$ denotes the Stratonovich product.
The work is the change in the energy due to changes in the control parameters $\bm{\lambda}$; since only the potential energy of the target system $U_{\bm{\lambda}}(\bm{x}^\text{s})$ depends on the control parameters, the work is independent of whether we describe the dynamics using \eqref{gle} or \eqref{langevin}.
The heat, on the other hand, is the change in energy due to the change in the configuration $\bm{z}$ of the extended system, which thus depends on the dynamics of both the target and auxiliary system.
Defining the entropy of the system as the differential entropy of the probability density $p_t(\bm{z})$, $S^\text{M}_t = - \Av{\ln p_t}_t$, we obtain the second law of thermodynamics \eqref{second-law} \cite{Sekimoto2010,Seifert2012}.
Using \eqref{first-law}, this can also be formulated as
\begin{align}
\dot{W}^\text{s}_t = \dot{\mathcal{F}}^\text{M}_t + T \sigma^\text{M}_t \geq \dot{\mathcal{F}}^\text{M}_t \label{second-law-work},
\end{align}
where the non-equilibrium free energy is defined as $\mathcal{F}^\text{M}_t = E^\text{M}_t - T S^\text{M}_t$.
While the work only depends on the target system, both the free energy and the entropy production depend on both the target and auxiliary system.

\paragraph{Appendix C: Information thermodynamics.}
In the information thermodynamics of bipartite Markov processes \cite{Horowitz2014}, the Markovian EPR \eqref{entropy-markov} is decomposed into two positive partial EPRs
\begin{gather}
\sigma_t^\text{M} = \sigma_t^\text{M,s} + \sigma_t^\text{M,a} \quad \text{with} \quad \sigma_t^\text{M,s} = \frac{1}{T} \Av{\bm{\nu}_t^\text{s} \cdot \bm{\gamma}^\text{s} \bm{\nu}_t^\text{s}}_t \label{partial-entropy}, 
\end{gather}
and similar for $\sigma_t^\text{M,a}$, where $\bm{\nu}_t^\text{s}(\bm{z})$ denotes the s-component of the vector field $\bm{\nu}_t(\bm{z})$.
Since $\bm{\nu}_t^\text{s}(\bm{z})$ depends on both $\bm{x}^\text{s}$ and $\bm{x}^\text{a}$, the partial EPR can only be defined in the extended system.
We also decompose the Markovian heat flow,
\begin{gather}
\dot{Q}^\text{M}_t = \dot{Q}_t^\text{M,s} + \dot{Q}_t^\text{M,a} \quad \text{with} \quad
\dot{Q}_t^\text{M,s} = \Av{\grad_{x^\text{s}} \mathcal{U}_{\bm{\lambda}} \circ \dot{\bm{x}}^\text{s}} , \label{heat-decomposition}
\end{gather}
and similar for $\dot{Q}_t^\text{M,a}$.
This is related to the change in the differential entropy of the target system $\dot{S}_t^\text{s}$ by the second law of information thermodynamics \cite{Sagawa2012,Ito2013,Hartich2014,Horowitz2014,Parrondo2015}
\begin{align}
\hat{\sigma}_t^\text{M,s} \equiv \dot{S}_t^\text{s} - \frac{1}{T} \dot{Q}_t^\text{M,s} = \dot{I}_t^\text{M,s} + \sigma_t^\text{M,s} \geq \dot{I}_t^\text{M,s} .
\end{align}
$\hat{\sigma}_t^\text{M,s}$ is the apparent EPR of the target system in the Markovian picture.
Just like the non-Markovian EPR $\hat{\sigma}_t^\text{NM}$, this quantity is not always positive; rather, it is bounded by the information flow $\dot{I}_t^\text{M,s}$ \cite{Horowitz2014},
\begin{align}
\dot{I}_t^\text{M,s} = \Av{\grad_{x^\text{s}} \ln p_t^{\text{a} \vert \text{s}} \cdot \bm{\nu}_t^\text{s}}_t \label{information-flow}.
\end{align}
The information flow decomposes the change in mutual information between the target and auxiliary system,
\begin{gather}
I^\text{M,s:a}_t = D_\text{KL}\big(p_t \Vert p_t^\text{s} p_t^\text{a} \big), \quad
\dot{I}_t^\text{M,s:a} = \dot{I}_t^\text{M,s} + \dot{I}_t^\text{M,a}  .
\end{gather}
The mutual information $I^\text{M,s:a}_t$ measures correlations between the target and auxiliary system; the information flow $\dot{I}_t^\text{M,s}$ quantifies the increase in correlations due to the dynamics of the target system.

From \eqref{partial-entropy}, it is obvious that $\sigma_t^\text{M,s}$ is invariant under invertible transformations of the auxiliary degrees of freedom, $\tilde{\bm{x}}^\text{a} = \bm{A} \bm{x}^\text{a}$, and neither does the differential entropy of the target system $\dot{S}_t^\text{s}$.
For linear transformations, the same is true for the information flow \eqref{information-flow}, as the conditional probability density of the auxiliary degrees of freedom only changes by the constant factor $\det \vert\bm{A}\vert$.
Writing \eqref{second-law-info} as
\begin{align}
\dot{Q}_t^\text{M,s} = T \big( \dot{S}_t^\text{s} - \dot{I}_t^\text{M,s} - \sigma_t^\text{M,s} \big),
\end{align}
$\dot{Q}_t^\text{M,s}$ is likewise invariant under linear variable transformations of the auxiliary degree of freedom and \eqref{interaction-heat} then implies the same for $\dot{Q}_t^\text{M,int}$.

\paragraph{Appendix D: Embedding and entropy production for linear and non-linear dynamics}
In the examples, we focus on the case where all forces in both the target and auxiliary system are linear.
Thus, the time-dependent statistics are Gaussian and can be completely characterized by the mean $\bm{\mu}_t = \av{\bm{z}(t)}$ and covariance matrix $\bm{\Xi}_t = \text{Cov}(\bm{z}(t))$ of the extended system, which evolve according to the ordinary differential equations
\begin{align}
    \dot{\bm{\mu}}_t &= - \bm{K}^\text{s}_{\bm{\lambda}(t)} \big(\bm{\mu}_t^\text{s} - \bm{a}_{\bm{\lambda}(t)}^\text{s} \big) - \bmc{K} \bm{\mu}_t \label{mean-cov-ode}\\
    \dot{\bm{\Xi}}_t &= - \big(\bm{K}_{\bm{\lambda}(t)}^\text{s} + \bmc{K} \big) \bm{\Xi}_t - \bm{\Xi}_t \big(\bm{K}_{\bm{\lambda}(t)}^\text{s} + \bmc{K} \big) + 2 \bm{\gamma}^{-1} T \n ,
\end{align}
corresponding to the potential $U_{\bm{\lambda}}(\bm{x}^\text{s}) = (\bm{x}^\text{s} - \bm{a}^\text{s}_{\bm{\lambda}})\cdot \bm{K}_{\bm{\lambda}}^\text{s} (\bm{x}^\text{s} - \bm{a}^\text{s}_{\bm{\lambda}})/2$.
All thermodynamic quantities such as heat, work, entropy and EPR, as well as the information flows, can be computed from the solution of these equations, see e.g.~Ref.~\cite{Ito2025} for explicit expressions.
To compute the data for Figs.~\ref{fig_trap_stiffness} and \ref{fig_power_law}, we integrate \eqref{mean-cov-ode} numerically.

While we assumed the dynamics of the auxiliary degrees of freedom in \eqref{langevin} to be linear, the target system can be subject to non-linear forces.
Then, the probability distribution of the extended system is no longer Gaussian, and \eqref{fpe} must be solved numerically.
If the dimension of the extended system is large, it is challenging to compute the probability distribution and often not feasible to compute information-theoretic quantities like differential entropy or information flows.
We stress that, despite this, energetic quantities such as work and heat \eqref{heat-work-markov} can be estimated directly from simulations of \eqref{langevin} by taking averages, making Markovian embedding useful even in these cases.
Moreover, newly developed methods such as machine-learning based approaches \cite{Kim2020,Otsubo2020} may be applied directly to the trajectories of the extended Markovian system to estimate quantities such entropy production.

\end{document}